\documentclass[review]{elsarticle}

\usepackage{lineno,hyperref}
    \usepackage{dingbat}
    \usepackage{amsmath}
        \usepackage{amssymb}

\journal{Journal of \LaTeX\ Templates}

\begin{document}

\begin{frontmatter}

\title{Understanding VSiPMT: a comparison with other large area hybrid photodetectors}

\author[UNINA,INFN]{F.C.T. Barbato\corref{mycorrespondingauthor}}
\cortext[mycorrespondingauthor]{Corresponding author}
\ead{barbato@na.infn.it}

\author[INFN]{G. Barbarino}

\address[UNINA]{Department of Physics, University of Naples Federico II}
\address[INFN]{Istituto Nazionale di Fisica Nucleare - Section of Naples}

\begin{abstract}
High energy physics community is continuously working on development of new photodetectors able to improve their experiments detection performances and so to increase the possibility of new discoveries and new important scientific results.\\
The first attempt to enhance the performances of classical PMTs by substituting the dynode chain is represented by MCP-PMTs. In addition to that, one of the strategies adopted to reach this goal is to realize hybrid photodetectors, that are new photomultipliers containing a semiconductor device within a vacuum tube.\\
In the last years, basically three competitors entered in this new class of photosensors: HPDs, ABALONE and VSiPMT.
In this article we will analyze how the operation principle of VSiPMT affects its detection performances and we will compare them to the other new photosensors: MCP-PMTs, HPDs and ABALONE.
\end{abstract}


\end{frontmatter}


\section{Introduction}
\begin{figure}[h!]
		\centering
		\includegraphics[scale=.5]{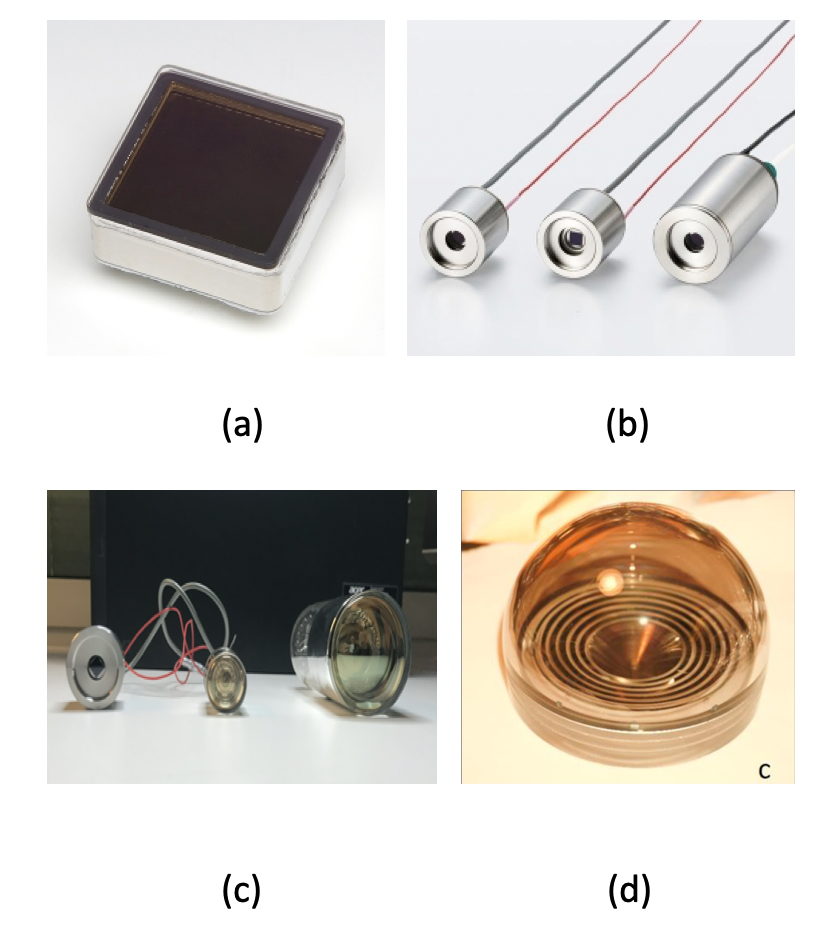}
		\caption{Pictures of the four new photodetectors: (a) MCP-PMT, (b) HPD, (c) VSiPMT, (d) ABALONE}\label{ibridi-foto}
	\end{figure}

A new generation photodetector is a device able to overcome the limits of the current generation.\\
Since more than one century, PMTs are quintessentially large area photons detectors for fundamental physics experiments. Despite their wide use in fundamental physics research experiments as well as in medical and industrial applications, they have several limitations all ascribable to their gain concept: the dynode chain, which provides the multiplication of secondary electrons.\\
Among them, some are due to physical behaviours linked to the device operation concept, like: significant fluctuations in the first dynode and anticorrelation between gain and linearity. Others, instead, are on a technical level, like: necessity for high voltage stabilization, power consumption, possible redioactive contamination during dynode chain manufacturing, complex and fragile mechanics.\\
High energy physics community and industries continuously work on the realization of new devices aimed at overcome this limits. In the years semiconductor based photosensors was a very fast growing sector that represents a strong answer to most of the above mentioned limitations. Nevertheless, semiconductor photosensors size is forced by the dark noise. As a consequence, for large detection areas a new strategy is necessary.  One of the adopted strategies is to realize hybrid photodetectors, that are obtained by the fusion of two technologies: vacuum tubes and solid state devices. 
Obviously, this cathegory evolves rapidly in parallel to semiconductors. As well as new solid state detectors are realized, new hybrids concepts become possible.\\
At the moment, new devices can count basically on four competitors: three hybrids (HPDs, ABALONE and VSiPMT) and MCP-PMTs, see FIG. \ref{ibridi-foto}, \citep{HPD, MCP, ABALONE, Barbarino2008}.
	
This devices are a clear example of how the hybrids concepts change in parallel to semiconductors evolution and we can classify them as \textit{before-SiPM} (HPDs) and \textit{post-SiPM} (VSiPMT and ABALONE). \\
All of them have been developed with the same goal: enhance the performances of the next future experiments. In the following sections we will analyze the detection features of the four competitors in detail and compare pros and cons of each solution.

\section{Operation principle and gain mechanism}
\begin{figure}[h!]
	\centering
		\includegraphics[scale=.3]{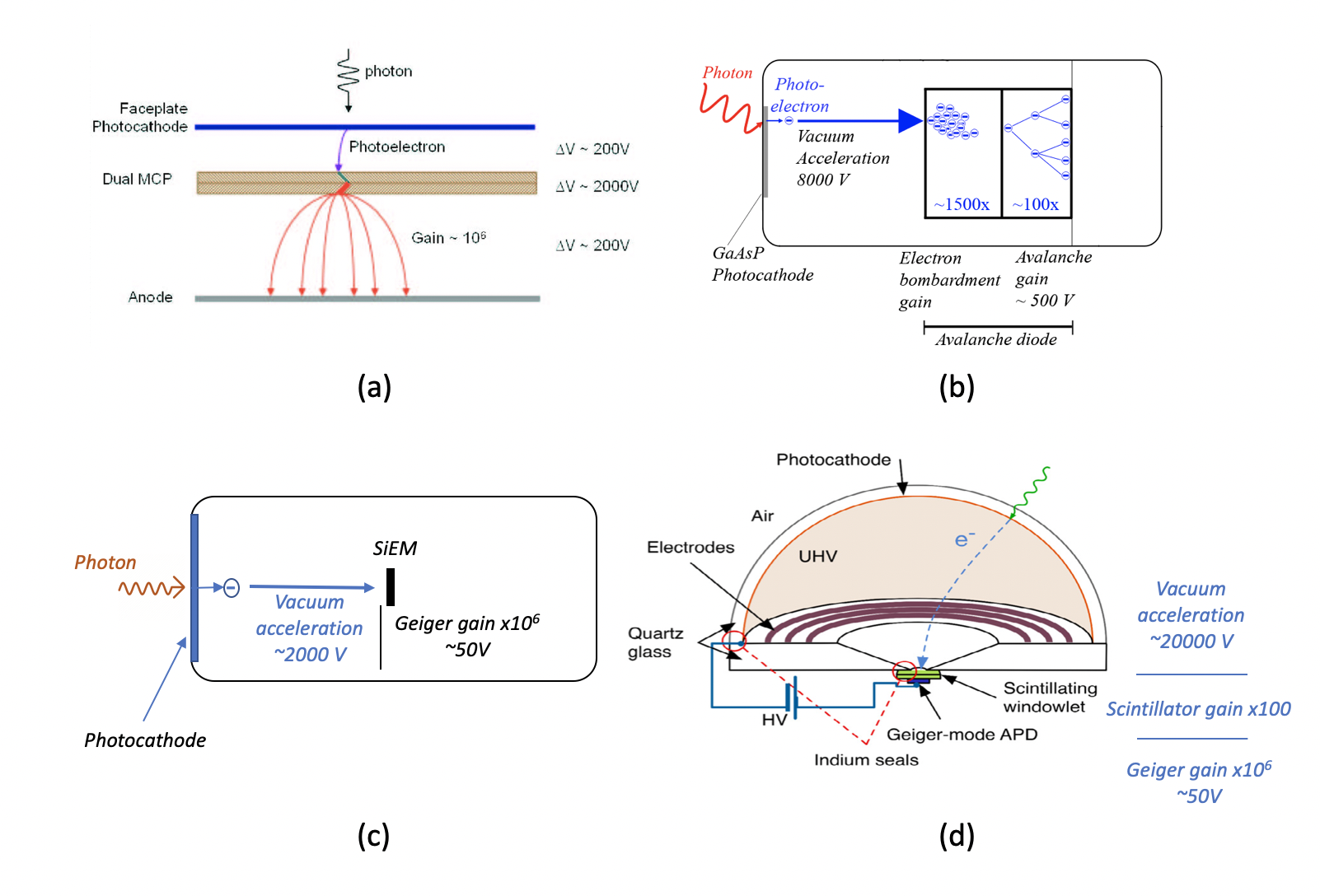}
		\caption{Concept designs of the four photodetectors: (a) MCP-PMT, (b) HPD, (c) VSiPMT, (d) ABALONE}\label{ibridi:concetti}
	\end{figure}

In order to clearly understand the solution to PMTs limits proposed by each device, it is necessary to resume quickly the operation principle they are based on and the gain concept they adopt.\\
All these four devices consists of a vacuum tube with a transparent front window where the photocathode is deposited. It converts light quanta into electrons. A focusing stage drives photoelectrons toward the amplification stage. At this point the four solutions split: MCP-PMTs and HPDs exploit the electron bombardment operation principle (i.e. the gain depends on the photoelectrons energy), VSiPMT exploits the geiger operation principle of the SiPM, while ABALONE combines the two operation principles, see FIG. \ref{ibridi:concetti}.
	
\paragraph{\textbf{MCP-PMT}} MCP-PMT exploits a double step of microchannel plates \citep{Microchannel} as amplification stage. The amplification concept is extremely simple, consisting of a two-dimensional array of a great number of glass capillaries tube (6-20$\mu m \varnothing$ ) whose inner surface acts as a secondary electron emitter. The relative position of the two MCPs must be chevron shaped as in FIG. \ref{ibridi:concetti}a to prevent ions-backscattering effects.\\
The photoelectron generated by the photocathode is accelerated towards the microchannels plate. The energy necessary for the first extraction of secondary electrons in the channels is quite low, $\sim 200 eV$. As for the dynodes chain, also inside the microchannels plate high voltage ($\sim 2kV$) is necessary to multiply secondary electrons, see FIG. \ref{ibridi:concetti}a. In this configuration a a gain of $\sim 10^6$ is reached.\\
It is evident that the MCP-PMT gain mechanism is nothing different from a multianode PMT, despite they use different materials for the extraction of secondary electrons.

\paragraph{\textbf{HPD}} HPDs, Hybrid Photon Diodes exploit silicon Avalanche PhotoDiodes (APDs) \citep{APD} as amplification stage.\\
In FIG. \ref{ibridi:concetti}b it is shown the basic concept of the device. A photon is converted to electron thanks to the photocathode. The photoelectron is accelerated towards the APD. Since APDs work in avalanche proportional mode, the higher is the photoelectron energy the higher is the number of electron-hole pairs produced in the APD, so the electron bombardment is necessary. The operation voltage of this device is thus approximately 8-10kV and the gain is $\sim 10^4$. HPDs thus need a sophisticated noise filter in order to correctly read the signal.

\paragraph{\textbf{VSiPMT}} VSiPMT, Vacuum Silicon PhotoMultiplier Tube, exploits a special Silicon PhotoMultiplier (SiPM) \citep{barbarino2011silicon}as amplification stage, see FIG. \ref{ibridi:concetti}c. This is here called Silicon Electron Multiplier (SiEM) since its structure is modified in order to detect even low energy electrons\footnote{The SiEM is a SiPM modified to detect electrons. Here, the epoxy resin layer is not present, the $SiO_2$ layer is thinner and the junction is p-n to allow electron penetration in the active region with a reasonable energy.} ($\sim 2 keV$) with high efficiency \citep{Barbato:astroparticle, Barbato:pollice, Barbato:2pollici}. In this configuration, the high voltage ($\sim 2kV$) is necessary to drive the photoelectron in the active region of the silicon bulk, so it can trigger the geiger avalanche. The geiger-mode operation principle ensures high gain, $\sim 10^6$, obtained in low voltage (the SiEM operation voltage is $\sim 50V$).

\paragraph{\textbf{ABALONE}} ABALONE, differently from VSiPMT, exploits a double amplification stage obtained by combining a standard SiPM with a scintillator layer, see FIG. \ref{ibridi:concetti}d. In this configuration, a photoelectron produced by the photocathode is accelerated toward the scinitillator windowlet. Here, a number of photons proportional to the photoelectron energy is produced and then they are detected by the coupled SiPM. The ABALONE inventors, designed the device to have a typical gain $\sim 10^8$. The geiger-avalanche of a SiPM gives a typical gain $\sim 10^6$. This value is combined with those obtained by the electron bombardment of the scintillator. Here a gain $\sim 100$ is obtained by operating the device with HV$\sim 20kV$.

\section{Fluctuations and photon counting}
In our opinion, the biggest limitation of PMTs is that they suffer of significant fluctuations. In a n-dynodes PMT, indeed, the number of secondary electrons, $\delta^n$, arriving at the anode during a single light pulse fluctuates. This happens because the energy necessary to extract an electron from the dynode $W_{dynode}\approx  30 eV$ and it depends on the ionization cross section, the mean free path inside the dynodes and the vacuum work function. Therefore, considering the typical n-dynodes PMT's supply voltage, the average number of secondary electrons extracted in each hit is $delta \approx  5$. This means that the spread in the output amplitude is dominated by fluctuations in the yield of the first dynode $\delta_{1st}$ where the absolute number of electrons is the smallest possible \citep{fluttuazioni}. 
Here, the standard deviation is $\sqrt{\delta_{1st}}$. Such a fluctuation affects the resolution of the single photon output pulse amplitude, that thus goes as:
	\begin{equation}\label{flutt}
		\propto \frac{\sqrt{\delta_{1st}}}{\delta_{1st}}.
	\end{equation} 
Since in PMTs, $\delta_{1st}$ is small, it is clear from eq. \ref{flutt} that these fluctuations are significants and makes impossible to separate cleanly an event caused by one photoelectron from one in which more photoelectrons are involved, resulting in a very poor photon counting capability.\\
In many scientific and medical applications, it is essential to detect low intensity signals down to a single photon and to count. Thus, it is important to analyze whether or not this limitation is still present in the four hybrids covered in this paper.\\

\begin{figure}[h!]
	\centering
		\includegraphics[scale=0.37]{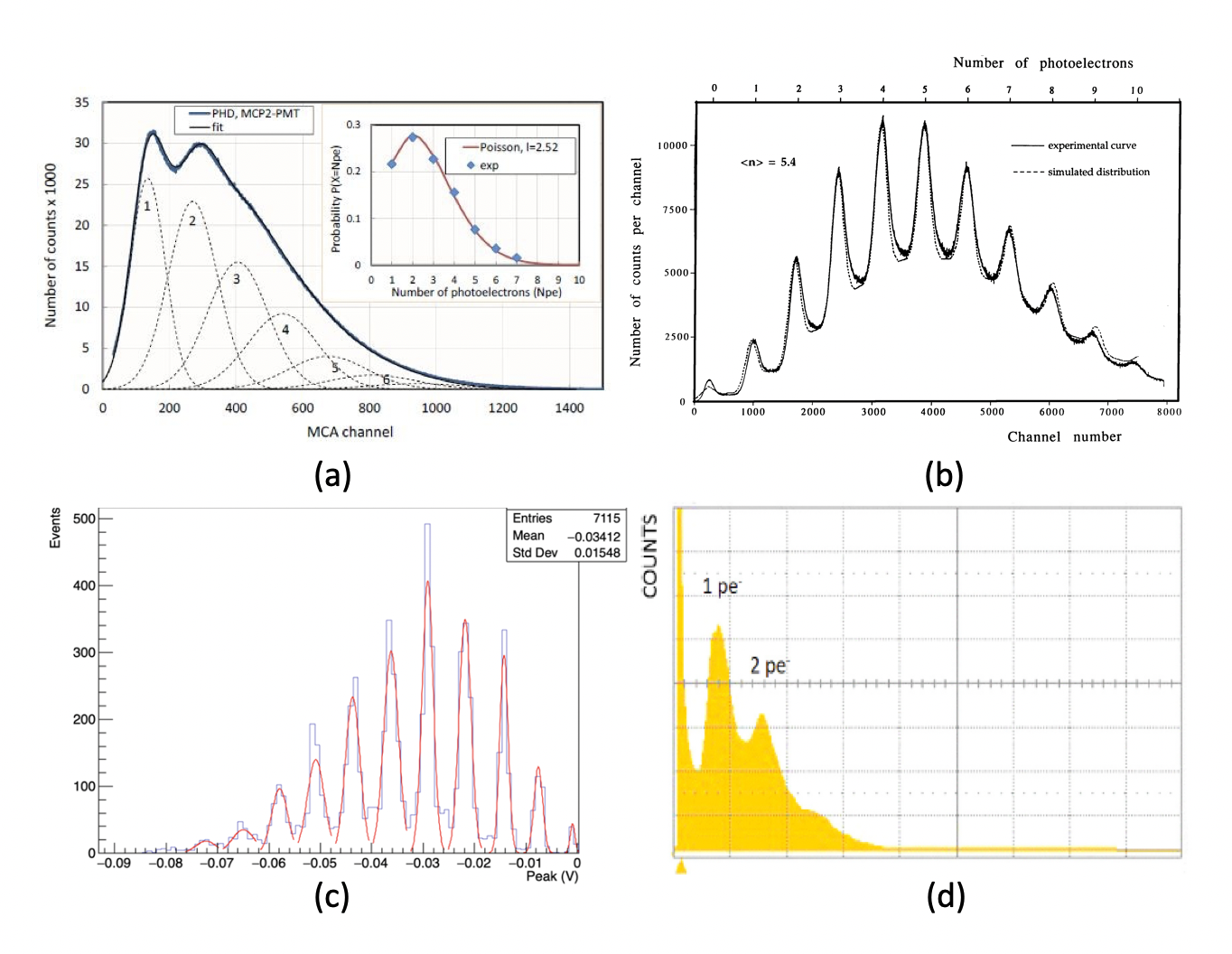}
		\caption{Pulse height distributions of the four devices: (a) MCP-PMT, Image credit: Photonis \citep{MCP}; (b) HPD, Image credit: C. Joram \citep{HPD}; (c) VSiPMT; (d) ABALONE, Image credit: D. Ferenc \citep{ABALONE}. }\label{fluttuazioni}
	\end{figure}

\textit{MCP-PMT} gain concept is very close to those of the dynode chain, see FIG. \ref{ibridi:concetti}b. It is given is given by the following equation using the length-to-diameter ratio of the channel
\begin{equation}
G_{MCP}= exp (\delta \cdot L/d)
\end{equation}
where $\delta$ is the secondary emission characteristics of the channel called gain factor. This gain factor is an inherent characteristic of the channel wall material and represented by a function of the electric field intensity inside the channel. The L/d ratio is designed in order to obtain $G\approx 10^3-10^4$ with HV = 1 kV. With a double stage of MCPs a very high gain is obtained. So, also in this case there is the production of secondary electrons in the microchannel. Same gain concept of PMTs but different means: even if this difference is very important in other aspects, e.g. the time response, it doesn't bring any improvement on the fluctuations problem. Also for this device, indeed, $\delta_{1st}$ is small, alike in PMTs, making photon counting difficult. As reported in \citep{photonis:mcp}, in MCP-PMTs it is possible in low illumination condition to separate the first and second peak. Nevertheless, the resolution of further peaks is not good enough because of the statistical increase of the pulse height distribution width, see FIG. \ref{fluttuazioni}a.\\

\textit{HPD} gain concept, instead, represents the first historical break. Here, indeed, the semiconductor in all its configurations (PIN, APD and SiPM) becomes an high sensitivity charge amplifier and the energy necessary to create an electron-hole pair falls to few electronvolts ($W_{e-h} \simeq 3.6 eV$), approximately ten times lower with respect to $W_{dynode}$, resulting in a equivalently high gain obtained in low voltage.\\
In this configuration, the accelerated photoelectrons bombard the silicon sensor and penetrate it to a depth of a few microns. Therefore, the gain of the device, is given by 
\begin{equation}
G_{HPD}=( HV - E_0)/W_{e-h}
\end{equation}
where, $E_o$ is the average energy lost, $\simeq 1 - 2keV$, in non-active silicon layers, while $W_{e-h}$ is the average energy necessary to create a single electron-hole pair in silicon, i.e. $\simeq 3.6 eV$, almost ten times lower than the secondary electrons extraction energy in dynodes. It is clear that here the electron bombardment, and so very high operation voltages, are necessary because the gain depends on $HV$ which should be $\sim 10 KV$ to ensure $G\approx 10^4$. Nevertheless, differently from PMTs, where large gain fluctuations are due to the Poisson distributed number of electrons after the first dynode, in HPDs, where $W_{e-h}$ is so low, the amplification process is obtained in a single stage and is totally dissipative. This gives gain variations that are much smaller to those of both a PMT and a MCP-PMT too and this is confirmed by the typical pulse height distribution of this kind of device \citep{fluttuazioni:HPD}, see FIG. \ref{fluttuazioni}b.\\

\textit{VSiPMT} gain concept goes even beyond HPD's. We can say that VSiPMT  projects SiPM's characteristics on a larger surface, i.e. the photocathode. From this point of view, the detection properties of the SiEM mounted inside the VSiPMT become exactely those of the VSiPMT itself. As a consequence, in this device, differently from all the others, the electron bombardment is not necessary. Here, indeed, the high voltage only drives photoelectrons towards the active layer of the SiEM. In this case, the gain is totally generated by the interaction of a photoelectron in silicon, where it generates a geiger-avalanche. Each microcell in the SiEM generates a highly uniform and quantized amount of charge every time an avalanche is generated  by  an  absorbed  photon  in  the  active  volume.  The  gain  of  a  microcell  (and  hence  the  sensor)  is  then  defined  as  the  ratio  of  the charge from an activated microcell to the charge on an electron. It can be expressed as follow
\begin{equation}
G_{VSiPMT}=\frac{C \cdot \Delta V}{q}
\end{equation}
As a consequence, each photon detected by VSiPMT results in a highly quantized output pulse with negligeable statistical fluctuations. In VSiPMT pulse height distribution, the peaks due to successive numbers of detected photons will be clearly visible. The separation between each pair of adjacent peaks is constant and corresponds to the charge generated from a  single fired microcell. With this mechanism, VSiPMT totally solve the fluctuations problem on large detection areas. This is confirmed by the typical pulse height distribution, alike the SiPM, where over than 10 photons are perfectly separated, see FIG. \ref{fluttuazioni}c.\\

\textit{ABALONE} gain is conceived in two phases. The first exploits electron bombardment of a scintillator windowlet. In this phase, obtained with HV = 23kV, a photon is converted by the photocathode to an electron which bombard the scintillator that produces $\approx 650$ photons in relation to its light yield (LY). In this way the single photon becomes 650 photons to be detected in the second gain step by the optically coupled SiPM. The overall gain of this device will be
\begin{equation}\label{gain_abalone}
G_{ABALONE}= LY_{scint}\cdot PDE_{SiPM} \cdot g_{SiPM} 
\end{equation}
If this guarantees a very high gain ($\approx 6\times10^8$) and thus the possibility to operate the device without amplifiers, on the other side the scintillator energy resolution seriously affects the absence of fluctuations typical of a SiPM. In addition, in this configuration further fluctuations can be induced by backscattering effects due to the high Z scintillator. From this viewpoint the ABALONE detector still suffer of fluctuations, confirmed by the pulse height distribution of the device, see FIG. \ref{fluttuazioni}d.

\section{Gain and linearity}
In PMTs gain and linearity are an inseparable duo. When gain increases a space charge effect arises in the last dynodes due to high current density. It can influence the electron trajectories, causing collection losses and at higher currents it can cause some electrons to return to the surfaces from which they originate. It is clear that this effect reduces the linearity range, since this condition is early reached in case of high intensity light pulses \citep{pmt_princ&app}.\\
As for the fluctuations problem, also in this case is important to analyze how the relation between gain and linearity is conditioned by the different gain mechanisms adopted in the devices under exam.\\

As already said, MCP-PMTs adopt the same gain mechanism of PMTs even if they exploit different materials. Once again, if on one side this design guarantees unequalled time response, on the other side it preserves all the limits ascribable to the gain mechanism itself. Therefore, with two MCPs the saturation occurs at the ends of the second MCP's channels \citep{linear:MCP}, after many electron multiplications and MCP-PMTs linearity is anticorrelated to the gain, alike it happens in PMTs.\\

In HPDs the gain is proportional to the photoelectron energy. Since the APD mounted inside the device has a low internal gain (i.e.$\approx 100$), most of the work is done by the front-end electronics. If on one side this entails the necessity of a highly performant electronics, on the other side it also means that the linearity of this device is not affected by the gain. So, HPDs overcome this PMT's limit.\\

In VSiPMT, as well as for SiPMs, the dynamic range and so the linearity depends only by a manufacturing factor: the number of pixels present on the SiEM mounted inside the device. Since for this design the gain is totally provided by the G-APD pixels, it means that gain and linearity are totally independent.\\

ABALONE, which uses a SiPM for the second amplification stage, obeys to the same law of VSiPMT and SiPM itself: the dynamic range and so the linearity only depends by the number of pixels present in the SiPM mounted. Nevertheless, one should remember that this design comes with a further amplification provided by the electron bombardment of a scintillator layer read by the SiPM, see eq. \ref{gain_abalone}. Therefore, the framework is slightly different by the VSiPMT one's. The single photon configuration is different for the two devices: in ABALONE, indeed, the gain is extreem and the single photon condition corresponds to $\approx 100$ photons detected by the SiPM, due to the first amplification stage given by the scintillator (i.e. $LY_{scint}$). It is easy to understand that even if the linearity is only linked to a SiPM manufacturing factor, the presence of a scintillator amplifies the light condition before the SiPM and therefore the linearity results to be anticorrelated to $LY_{scint}$.

\section{Time characteristics}
\begin{figure}[h!]
		\centering
		\includegraphics[scale=0.3]{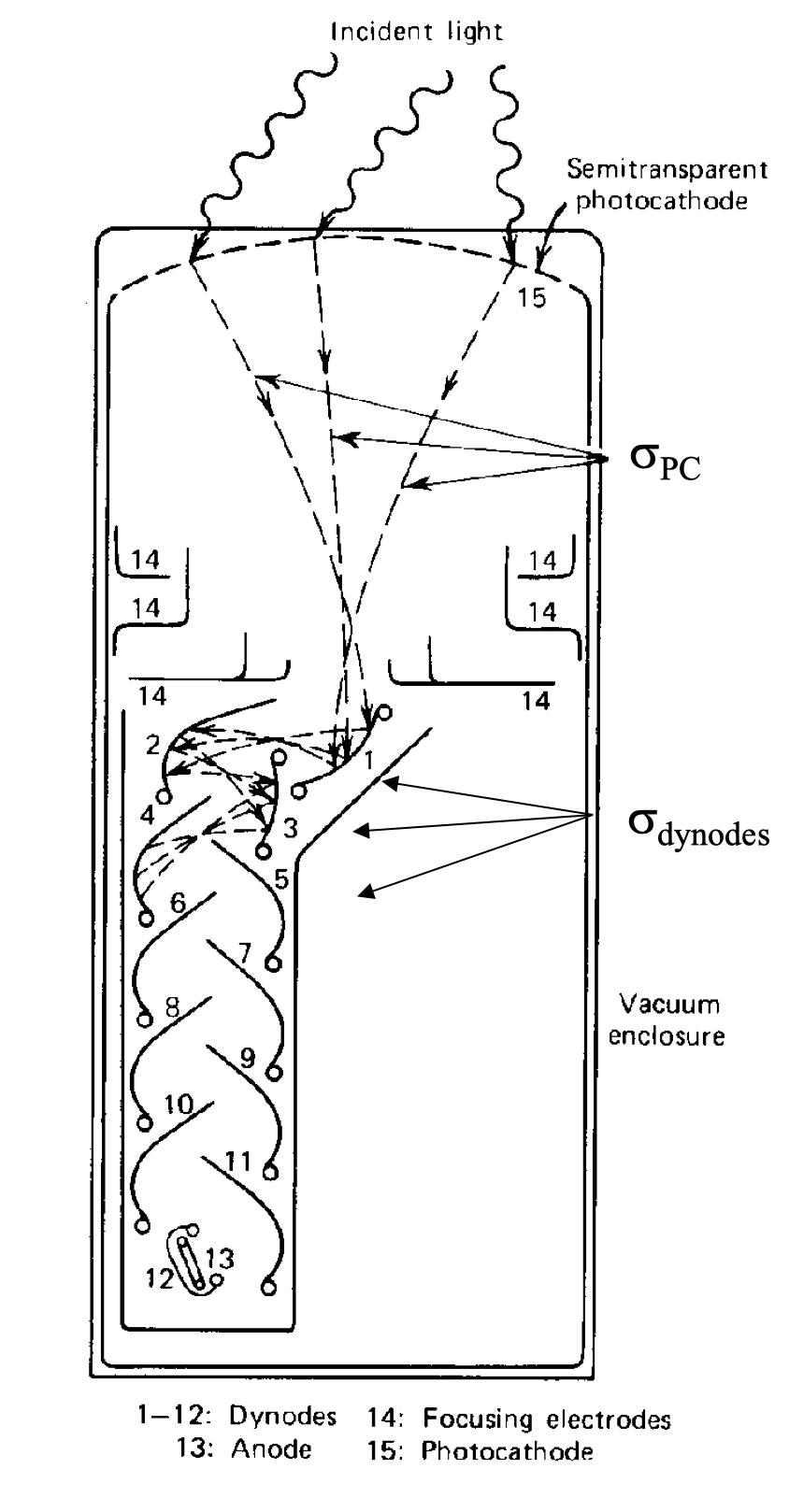}
		\caption{Scheme of the contribution to the time response of a PMT.}\label{fig:TTS}
	\end{figure}

There are two contributions to the time response of a PMT: the first is due to the different paths followed by photoelectrons going from the photocathode to the first dynode $\sigma_{PC}$, while the second is due to the different paths followed by the secondary electrons in the dynode chain $\sigma_{dynode}$, see FIG. \ref{fig:TTS}. The TTS can be therefore factorized as follow
\begin{equation}\label{TTS}
TTS = \sqrt{\sigma^2_{PC} + \sigma^2_{dynode}}
\end{equation}
Since we are dealing with large area PMTs, the condition $\sigma_{PC}\gg \sigma_{dynode}$  will always be valid and so $TTS \approx \sigma_{PC}$.
Among the four devices under exam, only one represents a real change from this point of view: the MCP-PMT. This device is realized following the so called proximity focusing design, in this configuration $\sigma_{PC}$ is minimized and so its time response is below 100 ps. \\
For all the others eq. \ref{TTS} is valid, in particular the second contribution is given by $\sigma_{APD}$ for the HPD and $\sigma_{SiPM}$ for both VSiPMT and ABALONE.\\
Photoelectron trajectories also conditions the photodetectors capability to work in strong magnetic fields. Since $\sigma_{PC}$ wheigh less for MCP-PMT, this device will convey less the magnetic field influence. Conversely, the others will be all influenced when operated in strong magnetic fields.

\section{Dark counts}
PMTs are very popular for their low dark noise. This excellent PMTs' characteristic in the transition to the hybrids get lost in some cases.\\
As already said, MCP-PMTs exploits the same gain principle of PMTs and it results in inherited its pros and cons, so MCP-PMTs maintain the same dark noise standard of PMTs.\\
Things changes when dealing with APDs and SiPMs. Dark noise in these two silicon devices, indeed, depends strongly both from the silicon bulk size and from is purity. Therefore, both HPDs and VSiPMT show a much higher dark noise with respect to PMTs according to the silicon device they use as amplification stage. Of course this characteristic can be improved following the solid-state technology progress or using some elctronics logic.\\
Even if also ABALONE exploits SiPM technology for the second amplification stage, in this case the situation is different from VSiPMT. As already said, due to the two amplification stages, the single photoelectron condition is here different and corresponds to $\approx 100$ photons read by the SiPM. If from one side this reduce the dynamic range of the ABALONE detector, on the other side it guarantees that a single photon event is registered with a signal that is 100 times higher with respect to the dark noise allowing the use of a threshold to cut the dark noise of the SiPM.

The outcomes due to physical behaviours analyzed up to now are resumed in table \ref{caratteristiche-fisiche}

\begin{table}[h!]
	\centering
	\includegraphics[scale=.35]{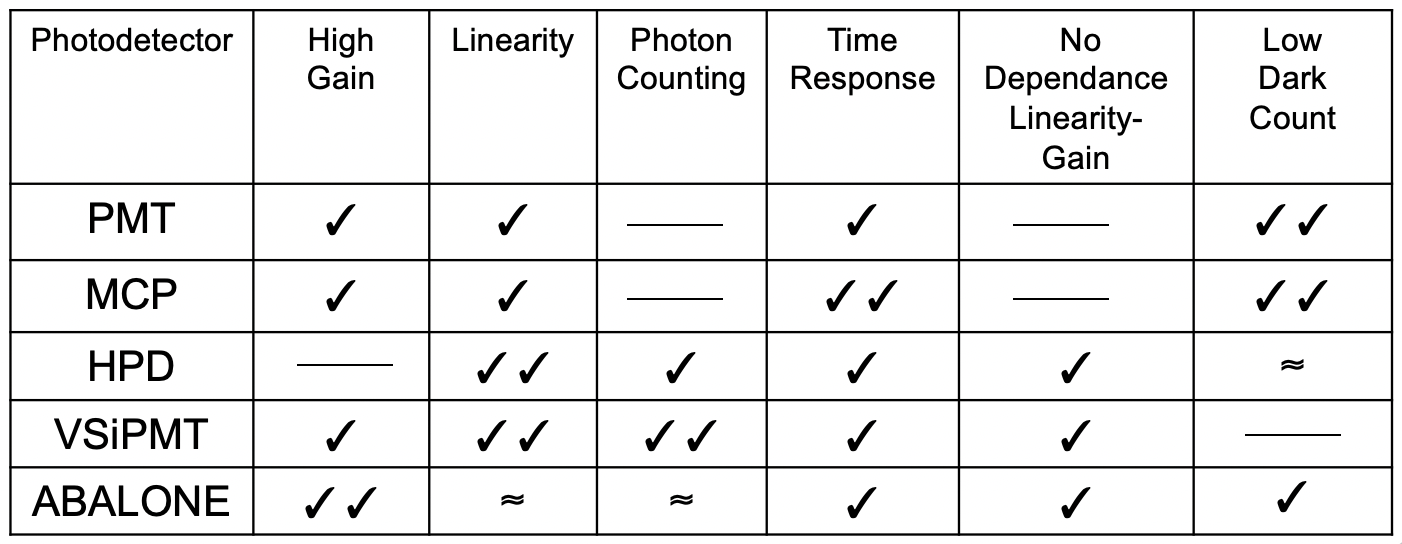}
	\caption{Resuming table of the outcomes due to physical behaviours. Legend:
	 \checkmark \checkmark  Fully satisfied;
	 \checkmark Satisfied;
	 $\approx$ Partially satisfied; 
	 --- Not satisfied}\label{caratteristiche-fisiche}
\end{table}

\section{Gain mechanisms outcomes on photodetectors' technical characteristics}
\begin{figure}[h!]
		\centering
		\includegraphics[scale=0.8]{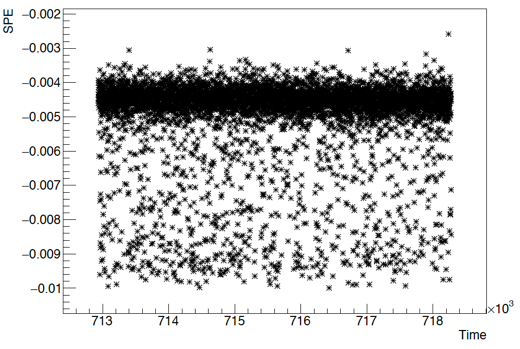}
		\caption{Measure of the VSiPMT SPE signal in a 85 minutes timeframe. During the measure the temperature was going from $0^\circ C$ to $65^\circ C$ with $5^\circ C$ step.}\label{temperatura}
	\end{figure}

The gain mechanism of a device has a visible effect also on its technical characteristics as: high voltage stability, power consumption, possible redioactive contamination during manufacturing, temperature dependence, complex and fragile mechanics.\\
The high voltage stability is necessary to guarantee gain stability during the acquisition process each time a device exploits electron bombardment. Therefore high voltage stability is a necessity for MCP-PMT, HPD and for ABALONE. On the contrary, this is not valid for VSiPMT which obtain high gain exploiting exclusively the geiger avalanche of the SiEM mounted inside, and therefore is the only device that amplifies the signal exclusively in low voltage.\\
With the introduction of hybrids, the power consumption problem has overcome. In PMTs it was, indeed, due to the presence of a voltage divider that was necessary to supply voltage to the dynode chain. \\
The absence of dynode chain also influences the presence of residual radioactivity. This characteristics results to be crucial for rare events experiments based on nobel liquids scintillators and depends on the materials used during the manufacturing. It is only slightly decreased in MCP-PMT, basically due to the glass present also in the amplification stage \citep{rad:MCP}. For all the others, exploiting the silicon devices as amplification stage the residual radioactivity can be limited. An HPD with extremely low radioactive background, below 1 mBq, was developed in past for dark matter research \citep{rad:HPD}. ABALONE, for his part, has been designed for low radioactivity experiments. VSiPMT level of radioactivity is expected to be low as well. Indeed, its mechanical composition is very similar to HPD's one and it is possible to suppose that also a similar level of radioactivity can be reached.\\
The temperature dependance of detection features is a key factor for all the experiments that experience substantial temperature variations during their data acquisition process (e.g. satellite experiments, telescopes, etc.). If this problem is absent in PMTs and MCP-PMTs, it is well known that gain in silicon detectors show a very strong temperature dependance. Neverthless, it has been measured that despite this strong dependance, when the silicon device is under vacuum conditions the thermal exchange with the external environment is very slow, see FIG. \ref{temperatura}.
However, also in this case is possible to use a temperature feedback supply voltage circuit to further stabilize the signal.\\
On the other side, in ABALONE, the SiPM is not under vacuum condition and therefore a power supply circuit with a temperature feedback is necessary.\\
Finally, the operation principle of these devices impinges also on their mechanical structure. If in PMTs the mechanical structure was very complex due to the presence of the dynodes chain and its voltage divider, this structure becomes simpler in hybrids. Among the four devices under exams, MCP-PMTs have the most complex mechanical structure because of the very fragile capillars of MCPs they mount inside. ABALONE, on the contrary, shows the simplest mechanical structure since the SiPM is not under vacuum conditions and so it doesn't need any feedthrough. In the average complexity level there are finally both HPDs and VSiPMT which using a single amplification stage resulting in a much more compact device with a strongly reduced number of feedthrough with respect to PMTs (only four output connections: HV, ground, APD or SiEM respectively anode and cathode).\\

The technical outcomes analyzed up to now are resumed in table \ref{caratteristiche-tecniche}

\begin{table}[h!]
	\centering
	\includegraphics[scale=.38]{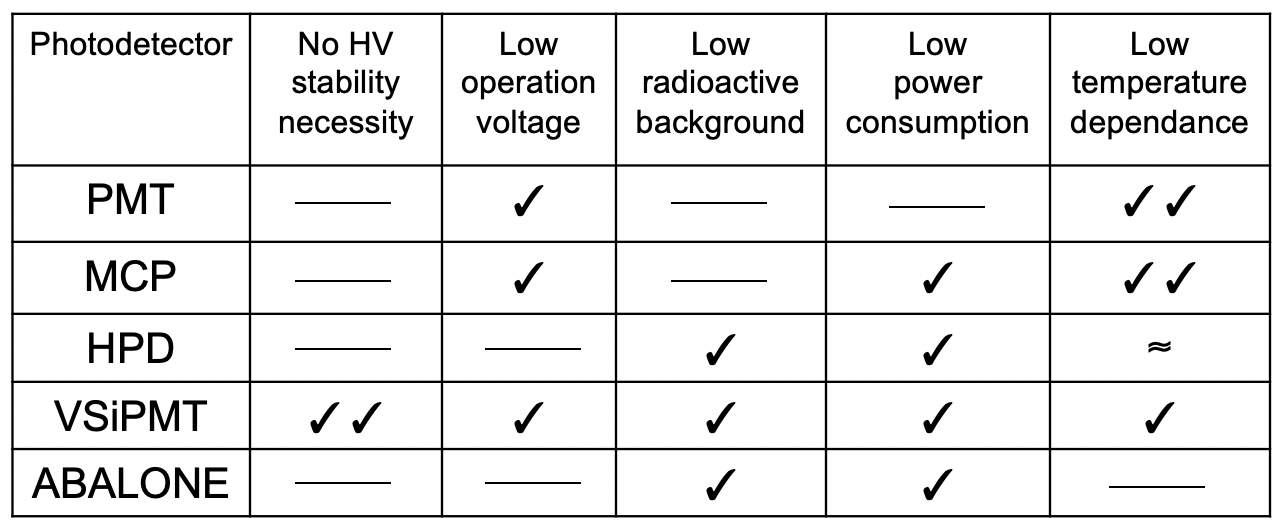}
	\caption{Resuming table of the technical outcomes. Legend:
	 \newline \checkmark \checkmark  Fully satisfied;
	 \checkmark Satisfied;
	 $\approx$ Partially satisfied; 
	 --- Not satisfied}\label{caratteristiche-tecniche}
\end{table}
	
\section{Conclusions}
Large area hybrid photodetectors are born with the intent to overcome the PMTs limits. Starting from this aim, in this article we analyzed how VSiPMT try to solve these issues in comparison to other new large area photodetectors, MCP-PMTs, HPDs and ABALONE. From this analysis it comes to light how all the characteristics examined are a direct consequence of the gain mechanism adopted by each device. Considering the VSiPMT design, it shows excellent features with respect to PMTs, such as photon counting capability, low power consumption and no linearity dependance from the gain, but at the same time it also shows a backward step in dark noise.\\
This analysis aims to define once and for all the differences in the detection features of the four photodetectors currently in the running. This comparison also wants to point out that there isn't a perfect device, but each of them excels in a field of application. For what concerns the VSiPMT the perfect fields of application are the Cherenkov telescopes and space experiments.

\section*{References}

\bibliography{mybibfile}

\begin{thebibliography}{10}
\expandafter\ifx\csname url\endcsname\relax
  \def\url#1{\texttt{#1}}\fi
\expandafter\ifx\csname urlprefix\endcsname\relax\def\urlprefix{URL }\fi
\expandafter\ifx\csname href\endcsname\relax
  \def\href#1#2{#2} \def\path#1{#1}\fi

\bibitem{HPD}
M.~{Suyama}, Y.~{Kawai}, S.~{Kimura}, N.~{Asakura}, K.~{Hirano}, Y.~{Hasegawa},
  T.~{Saito}, T.~{Morita}, M.~{Muramatsu}, K.~{Yamamoto}, A compact hybrid
  photodetector (hpd), IEEE Transactions on Nuclear Science 44~(3) (1997)
  985--989.
\newblock \href {http://dx.doi.org/10.1109/23.603790}
  {\path{doi:10.1109/23.603790}}.

\bibitem{MCP}
P.~Chen, J.~Tian, T.~Zhao, S.~Liu, S.~Qian, J.~Sun, S.~Si, H.~Liu, Y.~Wei,
  X.~Sai, L.~Guo, X.~Wang, Y.~Lu, C.~Wang, J.~Wang, K.~He, C.~Pei, L.~Tian,
  L.~Xin,
  \href{http://www.sciencedirect.com/science/article/pii/S0168900217311981}{Design
  of the large area mcp-pmt}, Nuclear Instruments and Methods in Physics
  Research Section A: Accelerators, Spectrometers, Detectors and Associated
  Equipment 912 (2018) 163 -- 166, new Developments In Photodetection 2017.
\newblock \href {http://dx.doi.org/https://doi.org/10.1016/j.nima.2017.11.012}
  {\path{doi:https://doi.org/10.1016/j.nima.2017.11.012}}.
\newline\urlprefix\url{http://www.sciencedirect.com/science/article/pii/S0168900217311981}

\bibitem{ABALONE}
D.~Ferenc, A.~Chang, C.~Saylor, S.~Böser, A.~D. Ferella, L.~Arazi, J.~R.
  Smith, M.~Å. Ferenc, {ABALONE$^{TM}$ Photosensors for the IceCube
  Experiment}, Nucl. Instrum. Meth. A954 (2020) 161498.
\newblock \href {http://arxiv.org/abs/1810.00280} {\path{arXiv:1810.00280}},
  \href {http://dx.doi.org/10.1016/j.nima.2018.10.176}
  {\path{doi:10.1016/j.nima.2018.10.176}}.

\bibitem{Barbarino2008}
G.~Barbarino, R.~de~Asmundis, G.~D. Rosa, G.~Fiorillo, V.~Gallo, S.~Russo,
  \href{http://www.sciencedirect.com/science/article/pii/S0168900208008462}{A
  new high-gain vacuum photomultiplier based upon the amplification of a
  geiger-mode p–n junction}, Nuclear Instruments and Methods in Physics
  Research Section A: Accelerators, Spectrometers, Detectors and Associated
  Equipment 594~(3) (2008) 326 -- 331.
\newblock \href {http://dx.doi.org/https://doi.org/10.1016/j.nima.2008.06.026}
  {\path{doi:https://doi.org/10.1016/j.nima.2008.06.026}}.
\newline\urlprefix\url{http://www.sciencedirect.com/science/article/pii/S0168900208008462}

\bibitem{Microchannel}
G.~Fraser,
  \href{http://www.sciencedirect.com/science/article/pii/S138738060100553X}{The
  ion detection efficiency of microchannel plates (mcps)}, International
  Journal of Mass Spectrometry 215~(1) (2002) 13 -- 30, detectors and the
  Measurement of Mass Spectra.
\newblock \href
  {http://dx.doi.org/https://doi.org/10.1016/S1387-3806(01)00553-X}
  {\path{doi:https://doi.org/10.1016/S1387-3806(01)00553-X}}.
\newline\urlprefix\url{http://www.sciencedirect.com/science/article/pii/S138738060100553X}

\bibitem{APD}
G.~Stillman, C.~Wolfe,
  \href{http://www.sciencedirect.com/science/article/pii/S0080878408601507}{Chapter
  5 avalanche photodiodes**this work was sponsored by the defense advanced
  research projects agency and by the department of the air force.}, in:
  R.~Willardson, A.~C. Beer (Eds.), Semiconductors and Semimetals, Vol.~12 of
  Semiconductors and Semimetals, Elsevier, 1977, pp. 291 -- 393.
\newblock \href
  {http://dx.doi.org/https://doi.org/10.1016/S0080-8784(08)60150-7}
  {\path{doi:https://doi.org/10.1016/S0080-8784(08)60150-7}}.
\newline\urlprefix\url{http://www.sciencedirect.com/science/article/pii/S0080878408601507}

\bibitem{barbarino2011silicon}
G.~Barbarino, R.~De~Asmundis, G.~De~Rosa, C.~M. Mollo, S.~Russo, D.~Vivolo,
  Silicon photo multipliers detectors operating in geiger regime: an unlimited
  device for future applications, in: Photodiodes-World Activities in 2011,
  InTechOpen, 2011.

\bibitem{Barbato:astroparticle}
G.~Barbarino, F.~Barbato, L.~Campajola, F.~Canfora, R.~[de Asmundis], G.~D.
  Rosa], F.~D. Capua], G.~Fiorillo, P.~Migliozzi, C.~Mollo, B.~Rossi,
  D.~Vivolo,
  \href{http://www.sciencedirect.com/science/article/pii/S0927650515000134}{A
  new generation photodetector for astroparticle physics: The vsipmt},
  Astroparticle Physics 67 (2015) 18 -- 25.
\newblock \href
  {http://dx.doi.org/https://doi.org/10.1016/j.astropartphys.2015.01.003}
  {\path{doi:https://doi.org/10.1016/j.astropartphys.2015.01.003}}.
\newline\urlprefix\url{http://www.sciencedirect.com/science/article/pii/S0927650515000134}

\bibitem{Barbato:pollice}
G.~{Barbarino}, F.~C.~T. {Barbato}, C.~M. {Mollo}, E.~{Nocerino}, D.~{Vivolo},
  A.~{Fukasawa}, {Another step towards photodetector innovation: The first
  1-inch industrial VSiPMT}, Astroparticle Physics 101 (2018) 27--35.
\newblock \href {http://dx.doi.org/10.1016/j.astropartphys.2018.01.001}
  {\path{doi:10.1016/j.astropartphys.2018.01.001}}.

\bibitem{Barbato:2pollici}
F.~Barbato, G.~Barbarino, G.~D. Rosa, R.~de~Asmundis, L.~Lavitola, C.~Mollo,
  M.~Mormile, D.~Vivolo, A.~Fukasawa,
  \href{http://www.sciencedirect.com/science/article/pii/S016890021930587X}{The
  2-inches vsipmt industrial prototypes}, Nuclear Instruments and Methods in
  Physics Research Section A: Accelerators, Spectrometers, Detectors and
  Associated Equipment 958 (2020) 162144, proceedings of the Vienna Conference
  on Instrumentation 2019.
\newblock \href {http://dx.doi.org/https://doi.org/10.1016/j.nima.2019.04.098}
  {\path{doi:https://doi.org/10.1016/j.nima.2019.04.098}}.
\newline\urlprefix\url{http://www.sciencedirect.com/science/article/pii/S016890021930587X}

\bibitem{fluttuazioni}
E.~I. Subbotin, F. M.;~Shubnikov, Fluctuations of the one-electron pmt current
  pulse, Radio Engineering and Electronic Physics 21 (1976) 148--152.

\bibitem{photonis:mcp}
P.~T. S.A.S., \href{https://www.azooptics.com/Article.aspx?ArticleID=11}{High
  quantum efficiency photocathodes}, AZoOptics.
\newline\urlprefix\url{https://www.azooptics.com/Article.aspx?ArticleID=11}

\bibitem{fluttuazioni:HPD}
C.~Joram, Large area hybrid photodiodes, Nucl. Phys. B, Proc. Suppl. 78 (1999)
  407--15.

\bibitem{pmt_princ&app}
S.-O. Flyckt, Photomultiplier tubes: principles and applications, Photonis,
  2002.

\bibitem{linear:MCP}
S.~{Matsuura}, S.~{Umebayashi}, C.~{Okuyama}, K.~{Oba}, Characteristics of the
  newly developed mcp and its assembly, IEEE Transactions on Nuclear Science
  32~(1) (1985) 350--354.

\bibitem{rad:MCP}
X.~Zhang, J.~Zhao, S.~Liu, S.~Niu, X.~Han, L.~Wen, J.~He, T.~Hu,
  \href{http://www.sciencedirect.com/science/article/pii/S016890021830593X}{Study
  on the large area mcp-pmt glass radioactivity reduction}, Nuclear Instruments
  and Methods in Physics Research Section A: Accelerators, Spectrometers,
  Detectors and Associated Equipment 898 (2018) 67 -- 71.
\newblock \href {http://dx.doi.org/https://doi.org/10.1016/j.nima.2018.05.008}
  {\path{doi:https://doi.org/10.1016/j.nima.2018.05.008}}.
\newline\urlprefix\url{http://www.sciencedirect.com/science/article/pii/S016890021830593X}

\bibitem{rad:HPD}
A.~Fukasawa, K.~Arisaka, H.~Wang, M.~Suyama,
  \href{http://www.sciencedirect.com/science/article/pii/S0168900210005036}{Qupid,
  a single photon sensor for extremely low radioactivity}, Nuclear Instruments
  and Methods in Physics Research Section A: Accelerators, Spectrometers,
  Detectors and Associated Equipment 623~(1) (2010) 270 -- 272, 1st
  International Conference on Technology and Instrumentation in Particle
  Physics.
\newblock \href {http://dx.doi.org/https://doi.org/10.1016/j.nima.2010.02.218}
  {\path{doi:https://doi.org/10.1016/j.nima.2010.02.218}}.
\newline\urlprefix\url{http://www.sciencedirect.com/science/article/pii/S0168900210005036}

\end{thebibliography}

\end{document}